\documentclass[preprint]{elsarticle}

\usepackage{epsfig,amssymb,latexsym}
\usepackage{rotating}
\usepackage{xcolor,url}
\usepackage{graphicx}
\usepackage{lscape}
\usepackage{amsmath,dsfont}
\usepackage{amsmath}
\usepackage{amssymb}
\usepackage{slashed}

\newlength{\MySep}
\newcommand {\bx}[1]  {\mbox{\boldmath $#1$}}

\newcommand {\beq}     {\begin{equation}}
\newcommand {\eeq}[1]  {\label{#1}\end{equation}}
\newcommand {\beqa}    {\begin{eqnarray}}
\newcommand {\eeqa}[1] {\label{#1}\end{eqnarray}}
\newcommand {\eeqan}   {\end{eqnarray}}

\makeatletter
\def\gsim{\compoundrel>\over\sim}
\def\lsim{\compoundrel<\over\sim}
\def\compoundrel#1\over#2{\mathpalette\compoundreL{{#1}\over{#2}}}
\def\compoundreL#1#2{\compoundREL#1#2}
\def\compoundREL#1#2\over#3{\mathrel
      {\vcenter{\hbox{$\m@th\buildrel{#1#2}\over{#1#3}$}}}}
\makeatother

\allowdisplaybreaks

\journal{Nuclear Physics A}

\begin{document}

\begin{frontmatter}



\title{Importance of chiral constraints for the pole content of the $\bar{K}N$ scattering amplitude}

\author[UJF]{P.~C.~Bruns}
\author[UJF]{A.~Ciepl\'{y} \corref{correspondence}}
\cortext[correspondence]{Corresponding author}
\ead{cieply@ujf.cas.cz}

\address[UJF]{Nuclear Physics Institute of the Czech Academy of Sciences, 250 68 \v{R}e\v{z}, Czech Republic}

\begin{abstract}
We critically examine the $\bar{K}N$ coupled-channel approach presented in \cite{Revai:2017isg} and 
demonstrate that it violates constraints imposed by chiral symmetry of QCD. The origin of this violation 
can be traced back to the off-shell treatment of the chiral-effective vertices, in combination 
with the use of non-relativistic approximations and the chosen regularization scheme. 
We propose an improved version of the approach, which is directly given by a resummation 
of relativistic Feynman graphs of baryon chiral perturbation theory, and is in accord with 
the chiral symmetry constraint. Within this improved model, two poles are generated dynamically 
in the isoscalar $\pi\Sigma - \bar{K}N$ coupled channels sector, in contrast with the non-relativistic 
model of \cite{Revai:2017isg} in which only one such pole was reported.
\end{abstract}


\end{frontmatter}

\section{Introduction}
\label{sec:intro}

The state-of-the-art description of the strangeness $S=-1$ meson-baryon scattering amplitude 
at low energies is provided by {\it chirally-motivated} coupled-channel calculations 
\cite{Kaiser:1995eg, Oset:1997it, Ikeda:2012au, Cieply:2011nq, Guo:2012vv, Mai:2014xna, Feijoo:2018den} 
based on solving an integral equation for the scattering amplitude $T$ with a kernel $V$ 
derived from baryon chiral perturbation theory (BChPT) \cite{Gasser:1987rb,Krause:1990xc,Bernard:2007zu}. 
The standard low-energy expansion 
of $T$ is not effective here due to the presence of the $\Lambda(1405)$ resonance just below 
the antikaon-nucleon threshold \cite{Kaiser:1995eg,Kaiser:2001hr}. A distinctive result of such 
approaches is the {\it two-pole structure} of the dynamically generated $\Lambda(1405)$ 
\cite{Oller:2000fj, Jido:2003cb, Hyodo:2011ur}. The use of non-perturbative extensions 
of the standard BChPT framework leads to a notable model-dependence in energy regions 
where the amplitude is not directly constrained by the existing experimental data, 
see e.g.~\cite{Cieply:2016jby}. In view of the importance of the antikaon-nucleon scattering 
amplitude for applications in few- and many-body calculations 
\cite{Shevchenko:2016wnu, Ohnishi:2017uni, Friedman:2007zza}, it is of great interest 
to assess the model-dependence of the coupled-channel amplitudes in the subthreshold region, 
and to find further theoretical and experimental constraints to reduce the ambiguities.

In a recent publication \cite{Revai:2017isg} (see also \cite{Revai:2019ipq}), J.~R\'{e}vai 
criticized a procedure commonly used 
to simplify the solution of the above-mentioned integral equations, namely, the so-called 
{\it on-shell factorisation} which reduces the integral equation to an algebraic equation that can 
be solved immediately. Solving a Lippmann-Schwinger equation (LSE) without applying 
this {\it approximation}, the author of \cite{Revai:2017isg} obtained a kernel without energy 
dependence (an advantageous feature for few-body calculations), and found that the solution 
``\ldots supports only one pole in the region of the $\Lambda(1405)$ resonance. Thus the almost overall 
accepted view, that chiral-based interactions lead to a two-pole structure of the $\Lambda(1405)$, 
becomes questionable'' \cite{Revai:2017isg}. In the following text we often refer to the work \cite{Revai:2017isg} 
as the JR approach.

In this contribution, we critically examine the model of \cite{Revai:2017isg} and the quoted 
conclusions concerning the $\Lambda(1405)$ structure. We will show that the model strongly violates 
certain general theoretical constraints derived from chiral symmetry, and thus spoils the motivation 
to employ a chiral-symmetric kernel from the outset. To make our criticism constructive, 
we devise a relativistic generalization of the model, henceforth referred to as the BC model, 
which can also be solved analytically without employing on-shell truncation. Although the resulting 
model is still not fully satisfactory in some aspects, we show that the pertinent generalized amplitude 
has an improved chiral behavior, and it brings back a second pole in the isoscalar $\pi\Sigma-\bar{K}N$ 
threshold region.

The outline of the paper is as follows. In the next section, we briefly review the JR approach presented 
in \cite{Revai:2017isg} and rewrite it in an equivalent and more transparent form that utilizes 
an effective potential with tadpole integrals accounting exactly for off-shell contributions 
to the meson-baryon loop function. The approach is improved in Section \ref{sec:BCmodel} 
by adhering to relativistic treatment of the effective potential and constraints arising 
from chiral symmetry. In Section \ref{sec:results} we provide fits of the model parameters 
to experimental data and compare the original JR model predictions with those made by the new 
model. The last section is reserved for concluding remarks.

\section{Discussion of the JR approach}
\label{sec:JRmodel}

Let us focus on the s-wave amplitude for antikaon-nucleon scattering. We shall mostly 
employ the notation of \cite{Revai:2017isg}, but leave out the channel indices, anticipating 
that the building blocks forming the coupled-channel scattering amplitude should thus be considered 
as matrices in the space of meson-baryon channels whenever appropriate. In particular, the on-shell 
scattering amplitude $T_{\mathrm{on}}$ introduced here is related to the commonly used relativistic 
s-wave amplitude $f_{0+}(\sqrt{s})$ \cite{Chew:1957zz} by
\begin{equation}
f_{0+}(\sqrt{s})\,\rightarrow\, -4\pi^2\sqrt{\mu}\: T_{\mathrm{on}}(k) \,\sqrt{\mu}
\label{eq:revaif0plus}
\end{equation}
in the non-relativistic limit. Here $\sqrt{s}$ denotes the total center-of-mass (c.m.) energy 
of the meson-baryon system, $\mu$ is a diagonal channel matrix containing the reduced masses 
$\mu_{j}$ of that system, and $k=\sqrt{2\mu(\sqrt{s}-m-M)}$, where $m$ stands for the (diagonal channel-matrix of) 
meson masses and $M$ for the (diagonal channel-matrix of) baryon masses. We note that momentum $k$ 
defined this way represents a non-relativistic approximation to the modulus of the c.m.~three-momentum 
$\bar{q}$ given as 
\begin{equation}
\bar{q} = \sqrt{(s-(M+m)^2)(s-(M-m)^2)}/(2\sqrt{s})  \: .
\label{eq:barq}
\end{equation}

From coupled-channel unitarity in the space of meson-baryon channels, we should always be able to write our 
{\it unitarized} amplitudes (with external particles on-shell) in the generic form
\begin{eqnarray}
  f_{0+}(\sqrt{s}) &=& -\left[K^{-1}_{\mathrm{rel}} + {\rm i}\bar{q}\right]^{-1} \,, \quad\mathrm{or}\quad    \nonumber \\ 
  4\pi^2\sqrt{\mu}\: T_{\mathrm{on}}(k)\, \sqrt{\mu} &=& \left[K^{-1} + {\rm i}k\right]^{-1} \:,
\label{eq:Kforms}
\end{eqnarray}
where the channel matrices $K_{\mathrm{rel}}$, $K$ are real in the physical region, and specify 
the chosen model parameterization. 
In \cite{Revai:2017isg}, a Lippmann-Schwinger equation (LSE) in the form 
\begin{equation}
T(\bx{p'},\bx{p};\sqrt{s}) = V(\bx{p'},\bx{p};\sqrt{s}) + \!\int\! 
                     d^3q\,V(\bx{p'},\bx{q};\sqrt{s})\frac{2\mu}{k^2-q^2+{\rm i}\epsilon}T(\bx{q},\bx{p};\sqrt{s})
\end{equation}
is solved for $T$ with a given kernel $V$. The on-shell s-wave scattering amplitude is then obtained by setting
the in- and outgoing meson momenta $p,\,p'$ on their mass shells, $T_{\mathrm{on}}(k)=T(\bx{p'},\bx{p};\sqrt{s})\bigr|_{p',p \rightarrow k}$. 
The debated {\it on-shell factorisation} corresponds to replacing the loop momentum $q$
in the argument of $T$ and $V$ by its on-shell value $k$, and subsequently pulling $T$ and $V$ 
out of the integral to arrive at an algebraic (matrix) equation. This procedure is {\it not} applied 
in \cite{Revai:2017isg}, and so the solution there is somewhat more elaborate. 

Following the notation used in \cite{Revai:2017isg}, we write the loop integrals occuring in the treatment of the LSE as
\begin{eqnarray}
  G_{AA} \!&=&\! 8\pi\mu\int_{0}^{\infty} \!dq\, \frac{q^2(u(q))^2}{k^2-q^2+{\rm i}\epsilon} \nonumber \\ 
         \!&=&\! -4\pi^2 \mu
            \left[\frac{\beta}{16}\left(5 \!-\! 15\left(k/\beta\right)^2 \!-\! 5\left(k/\beta\right)^4 
            \!-\!\left(k/\beta\right)^6\right) + {\rm i}k\right]
            (u(k))^2 \:, \label{eq:GAA} \\
  G_{AB} \!&=&\! G_{BA} = 8\pi\mu\int_{0}^{\infty} \!dq\, \frac{q^2(u(q))^2\gamma(q)}{k^2-q^2+{\rm i}\epsilon} 
             = \bar{\gamma}G_{AA} - I_{0} \:, \label{eq:GAB} \\
  G_{BB} \!&=&\! 8\pi\mu\int_{0}^{\infty} \!dq\, \frac{q^2(u(q))^2(\gamma(q))^2}{k^2-q^2+{\rm i}\epsilon} 
             = \bar{\gamma}^2G_{AA} -2\bar{\gamma}I_{0} - I_{1}  \:, 
\label{eq:GBB} 
\end{eqnarray}
where we introduced
\begin{equation}
  I_{n} := \frac{4\pi}{(2\mu)^{n}}\int_{0}^{\infty}\!dq\,q^2(u(q))^2(q^2-k^2)^{n}\: , 
\label{eq:In}
\end{equation}
for $n=0,1$, and
\begin{equation}
  \gamma(q) := \frac{q^2}{2\mu}+m\,,\quad \bar{\gamma} := \gamma(k)\,,\quad u(q) := \frac{\beta^4}{(\beta^2+q^2)^2} \: , 
\label{def:gammau}
\end{equation}
with $\beta$ standing for a cutoff parameter (often referred to as {\it inverse range}). 
Then, one can re-write the second equation in (\ref{eq:Kforms}) as
\begin{equation}
T_{\mathrm{on}}(k) = u(k)\left[\tilde{W}_{\rm JR}^{-1} - G_{AA}\right]^{-1}u(k) \: ,
\label{eq:TonW}
\end{equation}
where $\tilde{W}_{\rm JR}$ is again a real coupled-channel matrix depending only on $k^2$, and has no branch cuts. 
Therefore, it cannot contain the unitarity loop function $G_{AA}$. Indeed, we find that the on-shell s-wave 
amplitude derived in the JR approach can be brought into the form prescribed by Eq.~(\ref{eq:TonW}), 
when the {\it effective potential} reads
\begin{equation}
\tilde{W}_{\rm JR} = \left[\mathds{1}+\lambda I_{0}\right]^{-1}\left(\bar{\gamma}\lambda+\lambda\bar{\gamma} 
          - \lambda I_{1}\lambda\right)\left[\mathds{1}+I_{0}\lambda\right]^{-1} \:.
\label{eq:revaiW}
\end{equation}
Here $\mathds{1}$ denotes the unit matrix in channel space, and the coupling matrix $\lambda$ was specified 
in Sec.~2 of \cite{Revai:2017isg}. With the amplitude given by Eqs.~(\ref{eq:TonW}), (\ref{eq:revaiW}) and the appropriate parameter set, 
we reproduce exactly the resonance pole positions published in \cite{Revai:2017isg}. This way of rewriting 
the JR solution makes the relation to the {\it on-shell factorised} approach completely transparent: 
the latter is obtained by just dropping the terms containing the (diagonal channel matrices of) 
{\it tadpole integrals} $I_{n}$, a result that can be compared with Eq.~(18) in \cite{Revai:2017isg}. 
We also note the explicit results
\begin{equation}
I_{0} = \frac{\pi^2\beta^3}{8} \:, \qquad 
I_{1} = \frac{\pi^2\beta^3}{16\mu}(\beta^2-k^2)\: ,
\label{eq:I01num}
\end{equation}
obtained for the adopted form of $u(q)$, Eq.~(\ref{def:gammau}). Other off-shell extrapolations 
of the elementary vertices, or different regularisation schemes, would lead to different expressions 
for the functions $I_{n}$ appearing in the effective potential.

We note that the amplitude constructed above gives scattering lengths which do {\it not} 
vanish in the three-flavor chiral limit, as would be required by general arguments 
following from chiral symmetry and the (pseudo-)Goldstone-boson nature of the lowest meson octet. 
That is, even though a chiral-symmetric kernel is employed, the non-perturbative resummation 
framework spoils this attractive feature of the effective theory, inherited from QCD. To illustrate this, 
let us treat the one-channel case and calculate the corresponding scattering length derived 
from Eqs.~(\ref{eq:TonW}), (\ref{eq:revaiW}). At threshold, where $k=0$ and $\bar{\gamma}=m$, we obtain
\begin{equation}
a_{0+}^{\rm JR} = -4\pi^2\mu\left[\frac{(1+\lambda I_{0})^2}{2\lambda m -\lambda I_{1}\lambda}+4\pi^2\mu\frac{5\beta}{16}\right]^{-1} \:.
\label{eq:aJR}
\end{equation}
For the on-shell factorised solution with $I_{0},\,I_{1} \rightarrow 0$, one gets a prediction in line with 
the lowest-order ChPT\footnote{For the multichannel case, we can compare the s-wave scattering lengths 
following from the threshold limit of the amplitude $T_{\rm on}$ given by Eq.~(\ref{eq:TonW})
to the known results found in ChPT at the leading order, i.e.~neglecting the $\mathcal{O}(m^2)$
contributions, see e.g.~\cite{Kaiser:2001hr, Liu:2006xja, Mai:2009ce}:
\begin{equation*}
a_{0+,LO}^{\bar{K}N,I=0} = \frac{M_{N}}{4\pi(M_{N}+m_{K})}\frac{3m_{K}}{2F_{K}^2} \:, \quad  
a_{0+,LO}^{\bar{K}N,I=1} = \frac{M_{N}}{4\pi(M_{N}+m_{K})}\frac{m_{K}}{2F_{K}^2} \:.
\end{equation*}
}. 
However, if $I_{0,1} \not = 0$, Eq.~(\ref{eq:aJR}) provides $a_{0+}^{\mathrm{JR}}=4\pi^2mI_{1}/(I_{0})^2\,+\mathcal{O}(m^2)$. 
Treating $I_{1}$ as a quantity of order $\mu^{-1} \sim m^{-1}$, as suggested by Eq.~(\ref{eq:I01num}), 
we get an expression of order $\mathcal{O}(m^0)$, forbidden by ChPT. 
We presume here that the cutoff parameter $\beta$ stays fixed, or at least it does not vanish in the chiral limit. 
Therefore, the additional terms $\sim I_{n}$ which occur when the on-shell truncation is not performed
lead to a severe violation of chiral symmetry constraints.
From this observation, and similar ones 
for the multichannel case, we conclude that the effective potential given in Eq.~(\ref{eq:revaiW}) 
is not {\it soft} enough in the threshold region to be in accord with fundamental strictures 
dictated by chiral symmetry. One could argue that the chiral limit is a rather academic notion, 
not relevant for the physical amplitudes in question. However, as we already stated in the Introduction, 
theoretical constraints are badly needed in the present situation to reduce the model 
dependence. If one can construct an improved model, which is more reconcilable with chiral symmetry 
and does not introduce additional shortcomings, there is no reason to stick to the old model. 
In the following section, we attempt to devise an improved model with the same cutoff function $u(q)$ 
as used in \cite{Revai:2017isg}, but with a {\it softened} effective interaction kernel of $\mathcal{O}(m)$ 
at threshold. As we will see, obeying this chiral SU(3) constraint 
is quite relevant for the pole content of the amplitude.

\section{The hybrid BC approach}
\label{sec:BCmodel}

In an attempt to rectify some shortcomings of the JR approach we modify it by adhering 
to relativistic hadron kinematics while keeping the loop-integral regularization and not resorting 
to the {\it on-shell factorisation}. From the chiral Lagrangian at the leading order, 
one derives a vertex factor $-{\rm i}g(\slashed{q}_{i}+\slashed{q}_{j})$ for the meson-baryon 
scattering process $B(p_{a})M(q_{i}) \rightarrow B(p_{b})M(q_{j})$, where the $p_{a,b},q_{i,j}$ 
are off-shell four-momenta, and $g$ is a channel matrix of coupling constants that are related
to the $\lambda$ coefficients introduced in \cite{Revai:2017isg}. We multiply the vertex factor 
with the cutoff functions $u$ defined in Eq.~(\ref{def:gammau}) to obtain the new vertex factor
\begin{displaymath}
u(|\bx{q}_{j}|)\left(-{\rm i}g(\slashed{q}_{i}+\slashed{q}_{j})\right)u(|\bx{q}_{i}|)\, .
\end{displaymath}
The modification is not manifestly Lorentz-invariant, so from now on we shall work in the center-of-mass frame, 
where $p:=p_{a}+q_{i}=p_{b}+q_{j}=(\sqrt{s},\,\bx{0})$.
The summation of $s$-channel loop graphs (with $0,1,2\ldots$ loops) employing this vertex starts as
\begin{displaymath}
\hspace*{-50mm}
-{\rm i}\mathcal{T}(p_{a,b},q_{i,j}) = u(|\bx{q}_{j}|)\left(-{\rm i}g(\slashed{q}_{i}+\slashed{q}_{j})\right)u(|\bx{q}_{i}|) 
\end{displaymath}
\begin{displaymath}
\hspace*{10mm}
- {\rm i} u(|\bx{q}_{j}|) g \int\frac{d^4l}{(2\pi)^4}
\frac{(\slashed{q}_{j}+\slashed{l})\, {\rm i}(\slashed{p}-\slashed{l}+M)(u(|\bx{l}\,|))^{2}\, (\slashed{l}+\slashed{q}_{i})}
     {((p-l)^2-M^2)(l^2-m^2)} g u(|\bx{q}_{i}|) + \ldots
\end{displaymath}
It is possible to resum this infinite series in a closed form \cite{Borasoy:2007ku} (see also \cite{Nieves:2001wt, 
Lutz:2001yb, Djukanovic:2006xc, Bruns:2010sv, Morimatsu:2019wvk}). Setting the external momenta $p_{a,b},q_{i,j}$ 
on shell, the result is
\begin{equation}
 \mathcal{T}_{\mathrm{on}}(p_{a,b},q_{i,j}) = u(\bar{q})\left[\mathcal{W}^{-1}(\slashed{p}) - \mathcal{G}(\slashed{p},\beta)\right]^{-1}u(\bar{q})\,,
\label{eq:TonBC}
\end{equation}
where
\begin{eqnarray}
 \mathcal{W}(\slashed{p})     &=& \left[\mathds{1}+gI_{M}(\beta)\right]^{-1}\, \mathcal{W}_{0}(\slashed{p})\, 
                                  \left[\mathds{1}+I_{M}(\beta)g\right]^{-1}\: , \label{eq:newW} \\
 \mathcal{W}_{0}(\slashed{p}) &=& (\slashed{p}-M)g+g(\slashed{p}-M) + gI_{M}(\beta)(\slashed{p}-M)g \:, \\
 \mathcal{G}(\slashed{p},\beta) &=& \int\frac{d^4l}{(2\pi)^4}\frac{{\rm i}(u(|\bx{l}\,|))^2(\slashed{p}-\slashed{l}+M)}{((p-l)^2-M^2)(l^2-m^2)} \:, \\ 
 I_{M}(\beta)                 &=& \int\frac{d^4l}{(2\pi)^4}\frac{{\rm i}(u(|\bx{l}\,|))^2}{(l^2-m^2)} \:.
\end{eqnarray}
Explicit expressions for the loop functions $\mathcal{G}(\slashed{p},\beta)$ and $I_{M}(\beta)$ 
can be found in \ref{app:loopfunctions}. It is a standard procedure to compute the partial wave amplitudes of angular momentum 
$J=\ell\pm\frac{1}{2}$ from an invariant amplitude of the form $\mathcal{T}^{(1)}(s,z)\slashed{p}+\mathcal{T}^{(0)}(s,z)$, 
$z=\cos\theta_{\mathrm{cm}}$ \cite{Chew:1957zz}. We get
\begin{eqnarray}
  -16\pi\sqrt{s}f_{\ell\pm}(s) &=& \sqrt{E_{B}+M}\left(\sqrt{s}\mathcal{T}^{(1)}_{\ell}(s)+\mathcal{T}^{(0)}_{\ell}(s)\right)\sqrt{E_{B}+M} \nonumber \\ 
  &+&  \sqrt{E_{B}-M}\left(\sqrt{s}\mathcal{T}^{(1)}_{\ell\pm 1}(s)-\mathcal{T}^{(0)}_{\ell\pm 1}(s)\right)\sqrt{E_{B}-M} \:, 
\label{eq:fnullplus} 
\end{eqnarray}
where we used
\begin{displaymath}
  \mathcal{T}^{(0,1)}_{\ell}(s) = \int_{-1}^{1}dz\,P_{\ell}(z)\mathcal{T}^{(0,1)}(s,z)\,,\qquad E_{B} = (s+M^2-m^2)/(2\sqrt{s})\:.
\end{displaymath}

The minus sign on the l.h.s.~of Eq.~(\ref{eq:fnullplus}) stems from our phase convention for $\mathcal{T}$.
For $\mathcal{T}$ from Eq.~(\ref{eq:TonBC}), the angular integration is trivial, and we arrive at 
\begin{equation}
-8\pi\sqrt{s}f_{0+}(s) = \sqrt{E_{B}+M}u(\bar{q})\left[\mathcal{W}^{-1}(\sqrt{s}) - \mathcal{G}(\sqrt{s},\beta)\right]^{-1}u(\bar{q})\sqrt{E_{B}+M}\,.
\end{equation}
Essentially, the Dirac structure matrix $\slashed{p}$ is just replaced by $\sqrt{s}$ when we go from 
$\mathcal{T}_{\mathrm{on}}(p_{a,b},q_{i,j})$ to the partial wave amplitude $f_{0+}$. 
The expression for $f_{0+}$ can be cast in a form which is directly comparable to 
Eqs.~(\ref{eq:revaif0plus}), (\ref{eq:TonW}):  
\begin{eqnarray}
  f_{0+}^{\mathrm{BC}}(\sqrt{s}) &=& -4\pi^2\sqrt{\mu}\: T_{\mathrm{on}}^{\mathrm{BC}}(\sqrt{s})\, \sqrt{\mu}\, , \\
  T_{\mathrm{on}}^{\mathrm{BC}}(\sqrt{s}) &=& u(\bar{q})\left[\tilde{W}^{-1}_{\mathrm{BC}}(\sqrt{s}) 
                                              - G_{AA}^{\mathrm{rel}}(\sqrt{s})\right]^{-1}u(\bar{q})\: .
\end{eqnarray}
Here, within the {\it relativistic} BC approach, the effective potential and Green function integral read
\begin{eqnarray}
\tilde{W}_{\mathrm{BC}}(\sqrt{s}) \!\!&=&\!\! \sqrt{\frac{E_{B}+M}{\mu}}
    \frac{\mathcal{W}(\sqrt{s})}{4(2\pi)^3\sqrt{s}}\left[\mathds{1}+\left(I_{B}(\beta)-I_{M}(\beta)\right)
    \frac{\mathcal{W}(\sqrt{s})}{2\sqrt{s}}\right]^{-1}\!\!
    \sqrt{\frac{E_{B}+M}{\mu}}\, , \nonumber \\
  G_{AA}^{\mathrm{rel}}(\sqrt{s})  \!\!&=&\!\! 4(2\pi)^3\mu\sqrt{s}\:I_{MB}(s,\beta)\: . 
\label{eq:WBC}
\end{eqnarray}
Again, we refer to \ref{app:loopfunctions} for the algebraic forms of the relativistic loop integrals 
$I_{MB},\,I_{M},\,I_{B}$. Matching of the BC amplitude to the one generated by the JR approach,
Eq.~(\ref{eq:revaif0plus}), at tree level shows that we must have
\begin{equation}
g_{ij} = -\frac{c_{ij}}{4F_{i}F_{j}}\: ,
\label{eq:cij}
\end{equation}
where the coupling matrix $c_{ij}$ stems from the $\lambda_{ij}$ factors introduced in \cite{Revai:2017isg}. 
Eq.~(\ref{eq:cij}) represents a correct result as derived from the chiral Lagrangian. Specifically, we have 
$c_{\bar{K}N,\bar{K}N}^{I=0}=3$.

The BC amplitude constructed in this section has the advantage that the scattering lengths 
derived from it are indeed of $\mathcal{O}(m)$.
Explicitly, with two open meson-baryon channels, we find for the isoscalar $\bar{K}N$ scattering length 
\begin{align}
a_{0+,\mathrm{BC}}^{\bar{K}N,I=0} &= \frac{3m_{K}b_{0+}^{\bar{K}N,I=0}(m)}
                                          {8\pi F_{K}^2 \left(1+\frac{m_{K}}{M_{N}}\right)}\; , \;\;\;
                                          b_{0+}^{\bar{K}N,I=0}(m) = b_{0+}^{\bar{K}N,I=0}(0) + \mathcal{O}(m)\; ,
                                     \label{eq:aBC} \\ 
b_{0+}^{\bar{K}N,I=0}(0)          &= \frac{
                                  1\!-\!\frac{1}{8}\tilde{I}_{M}^{(0)}(\beta_{\pi\Sigma})
                                  \left(
                                    \frac{29}{2}\!-\!7\tilde{I}_{M}^{(0)}(\beta_{\pi\Sigma})
                                  \right)
                                  \!-\! \frac{3}{8}\tilde{I}_{M}^{(0)}(\beta_{\bar{K}N})
                                  \left(
                                    1\!-\!\frac{7}{8}\tilde{I}_{M}^{(0)}(\beta_{\pi\Sigma})
                                  \right)^2
                                  }{
                                  \left[1 -\tilde{I}_{M}^{(0)}(\beta_{\pi\Sigma})
                                    \left(1-\frac{21}{32}\tilde{I}_{M}^{(0)}(\beta_{\bar{K}N})\right) 
                                    -\frac{3}{4}\tilde{I}_{M}^{(0)}(\beta_{\bar{K}N})
                                  \right]^2 
                                  }\; , \nonumber
\end{align}
where we introduced the dimensionless quantity $\tilde{I}_{M}^{(0)}(\beta_j)=I_{M}^{(m=0)}(\beta_j)/F_{0}^2$ 
with the tadpole integrals $I_{M}(\beta_j)$  
and the meson decay constant $F_{0}$ both evaluated in the three-flavor chiral limit. 
The coefficient $b_{0+}^{\bar{K}N,I=0}(m)$ is of $\mathcal{O}(1)$ in the meson-mass expansion, 
and so the scattering length obviously vanishes in the chiral limit, in line with the strictures 
imposed by chiral symmetry. It may also be seen from Eq.~(\ref{eq:aBC}) that the tree-level result 
of ChPT (see footnote 1) is recovered when the tadpole corrections are dropped. This is 
in contrast with the JR approach where the $I_1$ term present in Eqs.~(\ref{eq:revaiW}) 
and (\ref{eq:aJR}) generates a zeroth-order term in the scattering lengths, stemming 
from the integral $G_{BB}$ in Eq.~(\ref{eq:GBB}). Indeed, the main problem of the cutoff 
regularization approach is most directly seen in Eqs.~(\ref{eq:GAA})-(\ref{eq:GBB}). 
There, the $\gamma(q)$ factors in the integrands of $G_{AB}$ and $G_{BB}$, which are 
formally small, do not lead to a suppression of those integrals with respect to $G_{AA}$. 
This is due to polynomial terms containing powers of the cutoff parameter $\beta$, 
which constitutes an additional mass scale that does not vanish in the chiral limit. 
While the BC approach improves the situation, as its effective potential $\tilde{W}_{\mathrm{BC}}$ 
is {\it softer} (proportional to \mbox{$\sqrt{s}-M$)} than the one of Eq.~(\ref{eq:revaiW}), an expansion 
of the threshold amplitudes in $m$ still does not exactly reproduce the $\mathcal{O}(m)$ results 
of ChPT, since the tadpole integral $I_{M}$ is non-vanishing in the chiral limit. Had one used dimensional 
regularization, the mesonic tadpole integrals would be $\sim m^2\log m$, and the corresponding 
effective potential would be even softer. It is well known \cite{Gasser:1987rb} that such power 
counting problems generally appear when one introduces additional mass scales (like baryon masses 
or cutoffs) in the low-energy effective theory.

\begin{figure}[h]
\centering
\includegraphics[width=0.6\textwidth]{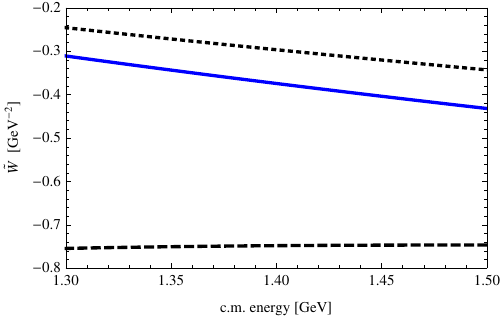}
\caption{$\bar{K}N$($I=0$) effective kernels over the c.m.~energy for the parameters of the model 
published in \cite{Revai:2017isg}. Black dotted - {\it pure} Weinberg-Tomozawa potential, 
black dashed - the $\tilde{W}_{\rm JR}$ kernel, blue continuous - the $\tilde{W}_{\mathrm{BC}}$ kernel.}
\label{fig:Wplots}
\end{figure}

The difference between the JR and BC approaches is demonstrated in Fig.~\ref{fig:Wplots}, where 
the pertinent isoscalar $\bar{K}N$ effective potential kernels $\tilde{W}$ are plotted in comparison 
with the classical Weinberg-Tomozawa (WT) kernel (tadpole integrals $I_0$ and $I_1$ set to zero). 
Apparently, the BC kernel is reasonably close to the WT one while the JR model is significantly 
off in the whole energy interval relevant for the $\pi\Sigma - \bar{K}N$ physics. In conclusion, 
the unfortunate combination of the off-shell extrapolation, the chosen regularization scheme 
and non-relativistic approximations employed in \cite{Revai:2017isg} leads to a strong departure 
from the leading chiral-symmetric kernel, forcing the {\it second pole} to move far away 
from the threshold region. The latter point may become more clear in the next section. 

\section{Results and discussion}
\label{sec:results}

The parameters of the chirally motivated approaches are standardly fixed in fits to the available 
low energy $K^{-}p$ experimental data that comprise the total cross sections 
\cite{Csejthey-Barth:1965izu, Sakitt:1965kh, Kim:1965zzd, Mast:1975pv, Bangerter:1980px, Ciborowski:1982et}, 
the threshold branching ratios $\gamma$, $R_c$ and $R_n$ \cite{Nowak:1978au, Tovee:1971ga} and the 1s level 
energy shift and absorption width due to the strong interaction in kaonic hydrogen \cite{Bazzi:2011zj}. 
In Ref.~\cite{Revai:2017isg} these observables were calculated treating the isoscalar and 
isovector sectors separately, adopting different inverse ranges $\beta_{MB}(I=0) \neq \beta_{MB}(I=1)$, 
though with physical masses used in the $\bar{K}N$ channels to ensure their correct threshold positions. 
As the latter isospin breaking leads to mixing of the $I=0$ and $I=1$ channels and the parameter space 
seems too large we adopt a more common approach and perform our calculations with proper physical 
particle masses and channels (comprising the $\pi\Lambda$, $\pi\Sigma$ and $\bar{K}N$ ones), 
and assuming $\beta_{MB}(I=0) = \beta_{MB}(I=1)$ as well. In addition, 
we constrain the fitted meson decay constants by fixing their mutual ratio adopting either $F_{K} = F_{\pi}$ 
or $F_{K} = 1.193\,F_{\pi}$, the second ratio complying with the PDG \cite{Rosner:2018pdg} 
as well as lattice results \cite{Aoki:2019cca} reviews. This leaves us with just 4 fitted parameters 
($F_{\pi}$ and three $\beta$'s) while 7 parameters were used in \cite{Revai:2017isg}.

The results of our fits are shown in Table~\ref{tab:fits} in comparison with the original 
JR model \cite{Revai:2017isg} and a version of the CS model \cite{Cieply:2011nq} restricted 
to the leading order WT interaction with only two fitted parameters, assuming $F_{K} = F_{\pi}$ and 
a common value $\beta_j = \beta_0$. As a different dataset and fitting approach was used 
in \cite{Cieply:2011nq} the $\chi^{2}/dof$ taken from this earlier publication cannot be compared 
directly with the $\chi^{2}/dof$ obtained for our current models. We also note that no $\chi^{2}/dof$ 
was provided in \cite{Revai:2017isg} but the same author presented $\chi^{2}/dof$ in more extendend 
fits performed later in \cite{Revai:2019ipq}. 

\begin{table}[h]
\caption{The fitted meson decay constants and inverse ranges, both in MeV, obtained in 
fits of low energy $K^{-}p$ data. Our results for models JR$_n$ and BC$_n$ are shown in comparison 
with earlier results provided by the CS and JR models. The label $n$ defines a fixed ratio 
of the meson decay constants, $F_{K}/F_{\pi} = 1.193^{n-1}$.}
\begin{center}
\begin{tabular}{c|cc|cc|ccc|c}
      &        &        & \multicolumn{2}{c|}{$I=0$ sector} & \multicolumn{3}{c|}{$I=1$ sector} & \\
model & $F_{\pi}$  & $F_{K}$  & $\beta_{\pi\Sigma}$ & $\beta_{\bar{K}N}$ 
      & $\beta_{\pi\Lambda}$ & $\beta_{\pi\Sigma}$ & $\beta_{\bar{K}N}$ & $\chi^{2}/dof$          \\ \hline
CS \cite{Cieply:2011nq} & 112.8  & 112.8  &  701.5  &  701.5  &  701.5  &  701.5  &  701.5 & 3.6  \\ 
JR \cite{Revai:2017isg} &  73.2  &  98.3  &  451.8  &  830.2  &  352.4  &  471.2  &  934.6 & ---  \\ 
JR$_1$                  & 116.3  & 116.3  &  553.2  &  860.6  &  656.3  &  553.2  &  860.6 & 2.62 \\ 
JR$_2$                  &  95.6  & 114.0  &  493.6  &  870.3  &  536.2  &  493.6  &  870.3 & 2.78 \\ 
BC$_1$                  & 105.9  & 105.9  &  876.7  & 1065.0  &  773.8  &  876.7  & 1065.0 & 2.39 \\ 
BC$_2$                  &  89.4  & 106.6  &  762.2  & 1125.8  &  637.8  &  762.2  & 1125.8 & 2.93 \\ \hline
\end{tabular}
\end{center}
\label{tab:fits}
\end{table}

The parameter values obtained in our fits are reasonable, in particular the meson decay constants 
of both BC models and of the JR$_2$ one are in agreement with predictions of other analyses 
\cite{Aoki:2019cca}, \cite{Tanabashi:2018oca}. In Table~\ref{tab:exp} we demonstrate how the considered 
models reproduce the $K^{-}p$ threshold observables. In general, our models tend to provide a bit 
too large values for the neutral channels branching ratio $R_n$ and for the kaonic hydrogen absorption 
width $\Gamma_{1s}$. With regards to the latter we find it appropriate to mention that in our 
fits (and in the $\chi^{2}/dof$ reported in Table~\ref{tab:fits}) we put more weight on the new 
kaonic hydrogen data and used smaller standard deviations of 20 and 50 eV for the energy shift $\Delta E_{1s}$ 
and the width $\Gamma_{1s}$, respectively, instead of the combined statistical and systematical errors 
reported in \cite{Bazzi:2011zj} and shown in Table~\ref{tab:exp}. In spite of this effort we were 
not able to bring completely the calculated $\Gamma_{1s}$ value within the experimental constraints. 
However, it is a fact that the description of the threshold observables (including the kaonic 
hydrogen ones) can be improved by accounting for the NLO contact terms in the chiral 
Lagrangian \cite{Ikeda:2012au}. These terms were completely disregarded in \cite{Revai:2017isg}, 
as well as in the JR$_n$ and BC$_n$ models introduced in the present work. 
On the other hand, the kaonic hydrogen characteristics predicted 
by the standard (utilizing {\it on-shell factorisation}) CS model \cite{Cieply:2011nq} are within the error bars despite the model 
is also restricted to the WT interaction kernel and its parameter space is quite narrow, 
just two parameters adjusted to the data.

\begin{table}[h]
\caption{Model predictions for the $K^{-}p$ threshold branching ratios $\gamma$, $R_c$, $R_n$ 
\cite{Nowak:1978au}, \cite{Tovee:1971ga}, and for the strong interaction energy shift $\Delta E_{1s}$ 
and absorption width $\Gamma_{1s}$ (both in eV) of the 1s level in kaonic hydrogen \cite{Bazzi:2011zj}.}
\begin{center}
\begin{tabular}{c|ccc|cc}
       & $\gamma$ &   $R_c$   &   $R_n$   & $\Delta E_{1s}$ & $\Gamma_{1s}$ \\ \hline
CS     &   2.36   &   0.636   &   0.183   &       329       &      643      \\ 
JR     &   2.35   &   0.687   &   0.203   &       384       &      462      \\ 
JR$_1$ &   2.36   &   0.647   &   0.214   &       261       &      717      \\
JR$_2$ &   2.36   &   0.649   &   0.226   &       279       &      709      \\ 
BC$_1$ &   2.35   &   0.638   &   0.215   &       285       &      666      \\ 
BC$_2$ &   2.35   &   0.642   &   0.223   &       294       &      669      \\ \hline
exp    &  2.36(4) & 0.664(11) & 0.189(15) &     283(42)     &    541(111)   \\ \hline
\end{tabular}
\end{center}
\label{tab:exp}
\end{table}

\begin{figure}[htb]
\centering
\includegraphics[width=0.4\textwidth]{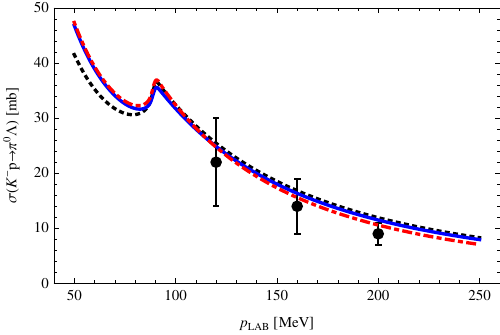} \hspace*{2mm}
\includegraphics[width=0.4\textwidth]{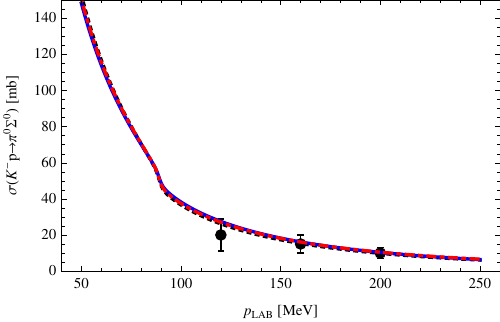} \\
\includegraphics[width=0.4\textwidth]{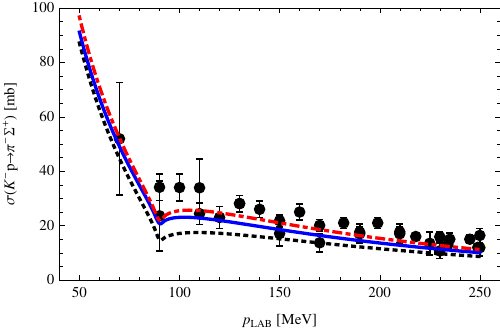} \hspace*{2mm}
\includegraphics[width=0.4\textwidth]{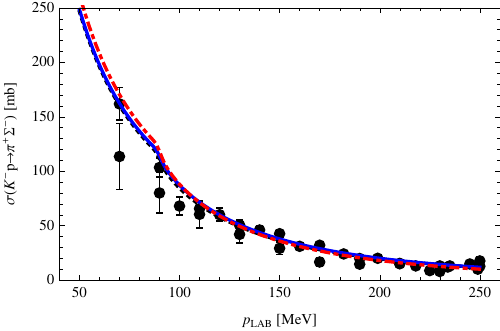} \\
\includegraphics[width=0.4\textwidth]{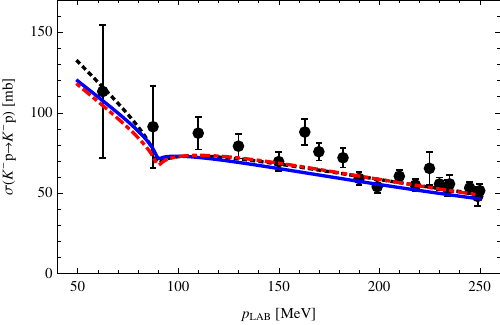} \hspace*{2mm}
\includegraphics[width=0.4\textwidth]{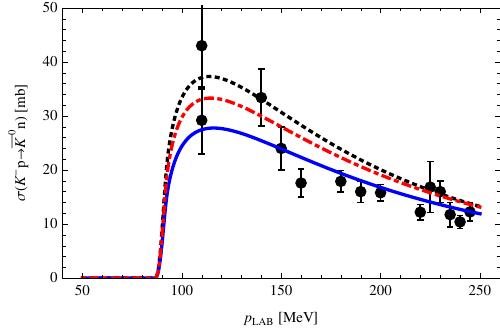} \\
\caption{Model predictions for the total cross sections for the $K^{-}p \rightarrow MB$ reactions.
Black dotted - CS model, red dot-dashed - JR$_1$ model, blue continuous - BC$_1$ model.}
\label{fig:xSect}
\end{figure}

The calculated cross sections are shown in Fig.~\ref{fig:xSect} as functions of the initial 
kaon momenta in the lab system. The upper limit of $p_{LAB} \lsim 250$ MeV is chosen to guarantee 
that we can safely neglect any p-wave contributions. Apparently, the models have no problem 
to reproduce the rather old bubble chamber data taken from Refs.~\cite{Csejthey-Barth:1965izu, 
Sakitt:1965kh, Kim:1965zzd, Mast:1975pv, Bangerter:1980px, Ciborowski:1982et}. In fact, the JR$_1$ 
and BC$_1$ model predictions are hard to distinguish in Fig.~\ref{fig:xSect}, the only exceptions 
being the $K^{-}p \rightarrow \bar{K}^{0}n$ and $\pi^{-}\Sigma^{+}$ cross sections. 
We also checked that the JR$_2$ and BC$_2$ models provide results completely akin to 
those provided by the models with $F_K = F_{\pi}$. In general, 
both the results presented in Table~\ref{tab:exp} and Fig.~\ref{fig:xSect} demonstrate 
that the experimental data are about equally well reproduced by all considered models.

Further, in Table \ref{tab:poles} we present the positions of the isoscalar and isovector poles 
generated by the models on the second Riemann sheet connected with the physical region by crossing 
the real axis in between the $\pi \Sigma$ and $\bar{K}N$ thresholds. The JR models provide only one isoscalar pole which appears 
to be their common feature. In Ref.~\cite{Revai:2017isg} the one-pole structure
is attributed to the fact that one does not resort to
on-shell factorisation when dealing with the loop function and the integration is performed over the whole 
domain of the intermediate meson-baryon off-shell momenta. However, our BC models generate two poles 
in the $I=0$ sector demonstrating that the {\it missing pole} is back when one improves the JR approach 
by keeping the relativistic form of the effective potential (proportional to $\sqrt{s}-M$). This result 
confirms the observation already stated in \cite{Morimatsu:2019wvk}: "The scattering T-matrix without 
on-shell factorization has two poles in the complex center-of-mass energy plane as with on-shell factorization ...". 
In fact, the two isoscalar poles also persisted in an earlier work on $\bar{K}N$ interactions \cite{Mai:2012dt} 
that treated the chiral expansion without resorting to on-shell factorisation of the loop function.
Of course, the $z_2$ pole position is very different from where one finds it when the on-shell factorisation 
is employed, as e.g.~in the CS model \cite{Cieply:2011nq}. While the LO+NLO approach presented 
in \cite{Mai:2012dt} has the $z_2$ pole shifted to higher energies, the pole is located much farther 
from the real axis in \cite{Morimatsu:2019wvk} and the BC models have it close to the $\pi\Sigma$ 
threshold, either very close to the real axis or even on it (and below the $\pi\Sigma$ threshold). 
The position of the pole, to which the $\pi\Sigma$ channel couples strongly, is apparently not 
sufficiently restricted by available experimental data. Thus its varied location within different approaches 
is not of big concern here. Just for completeness, in Table \ref{tab:poles} we also report 
an isovector $z_3$ pole found for all considered models at very varied locations, too far to affect 
physical observables at energies close to the $\bar{K}N$ threshold.

\begin{table}[h]
\caption{Pole positions (in MeV) on the [-,+] and [-,-,+] Riemann sheets for the $I=0$ and $I=1$ sectors, 
respectively.}
\begin{center}
\begin{tabular}{c|cc|c}
model   &   $z_1$ $(I=0)$  &  $z_2$ $(I=0)$  &  $z_3$ $(I=1)$    \\ \hline
CS      &  (1432.8, -24.9) & (1370.8, -54.2) & (1408.9,-199.7)   \\ 
JR      &  (1422.9, -25.7) &       ---       & (1106.5, -71.6)   \\ 
JR$_1$  &  (1442.8, -23.3) &       ---       & (1141.1, -80.5)   \\ 
JR$_2$  &  (1441.0, -22.5) &       ---       & (1266.4,   0.0)   \\ 
BC$_1$ &  (1439.9, -23.3) & (1316.0, -6.76) & (1361.1, -166.9)  \\ 
BC$_2$ &  (1437.8, -20.9) & (1251.1,  0.0)  & (1337.4, -117.3)  \\ \hline
\end{tabular}
\end{center}
\label{tab:poles}
\end{table}

We also checked the sensitivity of the $z_2$ pole position to the $\beta_{\pi\Sigma}$ value. This parameter 
is expected to have a significant impact on the appearance and location of the $z_2$ pole as 
the $\pi\Sigma$ channel coupling to the pole is large \cite{Cieply:2016jby}. For the BC$_1$ model, 
when the $\beta_{\pi\Sigma}$ parameter is reduced by 10\%, the pole shifts to (1326.6, -39.8) MeV, 
moving rather quickly away from the real axis. When the $\beta_{\pi\Sigma}$ value is increased by 10\%, 
the pole moves to the real axis, to the (1274.9, 0.0) MeV position. Such large variations confirm 
a strong sensitivity of the $z_2$ pole position to a particular setting of the BC model and indicate 
a possible model dependence in general.

\begin{figure}[htb]
\centering
\includegraphics[width=0.48\textwidth]{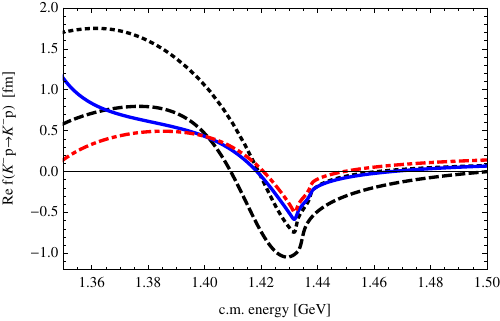}\hspace*{2mm}
\includegraphics[width=0.48\textwidth]{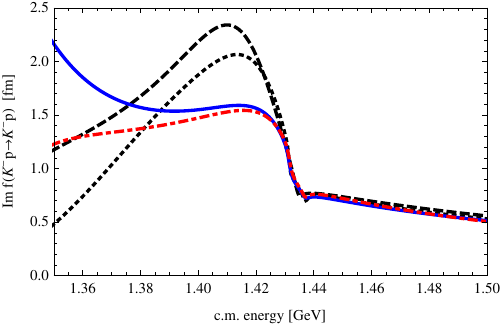} \\
\caption{Model predictions for the elastic $K^{-}p$ amplitude. The real (left panel) 
and imaginary (right panel) parts of the amplitude generated by our models BC$_1$ (blue continuous line), 
and JR$_1$ (red dot-dashed line) are shown in comparison with the CS model \cite{Cieply:2011nq} 
and JR model \cite{Revai:2017isg} predictions visualized by the black dotted and dashed lines, respectively.}
\label{fig:KpAmpl}
\end{figure}

To complete our discussion of the model predictions we look at the energy dependence of the elastic $K^{-}p$ 
amplitudes generated by the considered models and shown in Fig.~\ref{fig:KpAmpl}. There, we have opted for 
not presenting our models with the $F_K = 1.193\,F_{\pi}$ setting as their inclusion would only make 
the figure more cluttered. It is remarkable that our 
JR$_1$ and BC$_1$ models provide practically the same $K^{-}p$ amplitudes for energies $\sqrt{s} \gsim 1.4$ GeV. 
From about 1.42 GeV (1.43 GeV) these models are also in nice agreement with the CS model predictions 
for the real (imaginary) part of the amplitude. This is in accord 
with observations made in \cite{Cieply:2016jby} that the experimental data provide sufficient restrictions 
to determine the $K^{-}p$ amplitude around and above the channel threshold irrespective of the adopted theoretical 
approach. We note that the real part of the $K^{-}p$ amplitude generated by the original JR model differs 
even at these energies, most likely due to the different approach to the isospin breaking adopted in \cite{Revai:2017isg}. 
The differences between various models increase as one goes further below the $K^{-}p$ threshold into the sector 
not restricted by experimental data. We do not show the amplitudes below 1.35 GeV in Fig.~\ref{fig:KpAmpl} 
as the poles emerging close to (and below) the $\pi\Sigma$ threshold have huge impact on the amplitudes 
generated by some models. In particular, we have noted that the JR, JR$_1$ and BC$_2$ models also suffer 
from emergence of unphysical poles at various Riemann sheets (including the physical one) for energies below 
the $\pi\Sigma$ threshold. As these energies are too far below the $K^{-}p$ threshold, the emergence of such 
spurious states have no impact on physical observables used in our dataset. At the same time, it is difficult to eliminate 
solutions with very distant poles (i.e.~too far from any experimental data points) from the $\chi^{2}$ fits. 
Though, it is worth noting that the more established CS model, that utilises the {\it on-shell factorisation}, 
behaves quite regularly even at energies below the $\pi\Sigma$ threshold.

\section{Conclusions}
\label{sec:concl}

It is always good to scrutinize ad hoc procedures like the {\it on-shell factorisation}, 
and to question conclusions based on their use. In this respect, the study of Ref.~\cite{Revai:2017isg} 
is well justified. However, as we point out here, some care has to be taken in the construction
of the {\it chiral-unitarized} model amplitudes, in order not to violate basic constraints 
from the chiral symmetry of QCD. This symmetry is an important guiding principle 
of low-energy hadron physics, and yields the motivation to employ kernels derived from 
chiral-symmetric effective Lagrangians in the model amplitudes. As we have shown, the JR approach 
of \cite{Revai:2017isg} features an effective interaction which deviates strongly from the one 
derived from the leading chiral Lagrangian, and produces scattering lengths which are not suppressed 
by powers of the meson masses over the hadronic scale $\sim 1\,\mathrm{GeV}$. This happens just due 
to the inclusion of off-shell effects, which were meant to {\it improve} the theoretical description. 
We have provided a way out of this dilemma by explicitly constructing a refined version of the approach, 
employing relativistic propagators and Feynman rules directly corresponding to the leading chiral 
Lagrangian, while also avoiding the on-shell truncation. The effective kernel of this new version is
much closer to the leading chiral interaction (see Fig.~\ref{fig:Wplots}), and leaves us with 
scattering lengths that vanish in the three-flavor chiral limit. Performing simple fits to experimental 
data with both versions of the model, we have demonstrated that the {\it chirally improved} 
version yields two $I=0$ poles located on the second Riemann sheet in the energy region 
of the coupled $\pi\Sigma$, $\bar{K}N$ scattering processes, whereas in the non-relativistic 
model of \cite{Revai:2017isg}, only one pole is found in the relevant region of the complex-energy 
surface. While we have used here only the leading term in the chiral expansion of the kernel 
(Weinberg-Tomozawa interaction) to make a fair comparison with Ref.~\cite{Revai:2017isg}, 
we believe that the description can be further improved by the inclusion 
of higher-order interaction kernels, and that the inherent model-dependence can be constrained 
in a more extended analysis of additional experimental data, e.g.~the two-meson photoproduction 
data from CLAS \cite{Moriya:2013eb}. Some work in this direction was already done 
in \cite{Roca:2013av, Nakamura:2013boa, Mai:2014xna}.

\section*{Acknowledgement}

We thank J.~Mare\v{s} for careful reading of the manuscript, encouragement and comments. 
This work was supported by the Czech Science Foundation GACR grant 19-19640S.

\appendix

\section{Explicit expressions for loop functions}
\label{app:loopfunctions}
\def\theequation{\Alph{section}.\arabic{equation}}
\setcounter{equation}{0}

The fundamental loop integrals are defined as follows,
\begin{eqnarray}
I_{MB}(s,\beta) &=& \int\frac{d^4l}{(2\pi)^4}\frac{{\rm i}\,(u(|\bx{l}\,|))^2}{((p-l)^2-M^2)(l^2-m^2)}\biggr|_{p=(\sqrt{s},\bx{0})}\, , \\
I_{M}(\beta)    &=& \int\frac{d^4l}{(2\pi)^4}\frac{{\rm i}\,(u(|\bx{l}\,|))^2}{(l^2-m^2)}\, , \\
I_{B}(\beta)    &=& \int\frac{d^4l}{(2\pi)^4}\frac{{\rm i}\,(u(|\bx{l}\,|))^2}{(l^2-M^2)}\, . 
\end{eqnarray}
Explicitly, we get
\begin{equation}
  \!I_{MB}(s,\beta) \!=\! \frac{(u(\bar{q}))^2}{16\pi^2\sqrt{s}}
  \left[E_{B}h_{B}(\bar{q},\beta) \!+\! E_{M}h_{M}(\bar{q},\beta) \!-\! 4\bar{q}\,\mathrm{Artanh}\left(\frac{2\sqrt{s}\,\bar{q}}{(M+m)^2-s}\right)\right],
\label{eq:Imb}
\end{equation}
where the baryon and meson c.m.~energies are
\begin{displaymath}
  E_{B} = (s+M^2-m^2)/(2\sqrt{s})\: , \qquad E_{M} = (s-M^2+m^2)/(2\sqrt{s})\: ,
\end{displaymath}
and the functions appearing in Eq.~(\ref{eq:Imb}) read as
\begin{eqnarray}
  h_{B}(\bar{q},\beta) &=& \frac{h_{B}^{(0)}(\bar{q},\beta)}{24\beta^4(\beta^2-M^2)^3} + \frac{h_{B}^{(1)}(\bar{q},\beta)}{8\beta^5(\beta^2-M^2)^{\frac{7}{2}}}\,\mathrm{Artanh}\left(\frac{\sqrt{\beta^2-M^2}}{\beta}\right)\,,\\
  h_{B}^{(0)}(\bar{q},\beta) &=& \beta^6(44\beta^4-44\beta^2M^2+15M^4) \nonumber \\ &+& 3\beta^4(24\beta^4-10\beta^2M^2+M^4)\bar{q}^2 + 3\beta^2(12\beta^4+8\beta^2M^2-5M^4)\bar{q}^4 \nonumber \\ &+& (8\beta^4+10\beta^2M^2-3M^4)\bar{q}^6\,,\\
  h_{B}^{(1)}(\bar{q},\beta) &=& \beta^6(-16\beta^6+24\beta^4M^2-18\beta^2M^4+5M^6) \nonumber \\ &-& 3\beta^4M^2(16\beta^4-16\beta^2M^2+5M^4)\bar{q}^2 - \beta^2M^2(32\beta^4-22\beta^2M^2+5M^4)\bar{q}^4 \nonumber \\ &-& M^2(8\beta^4-4\beta^2M^2+M^4)\bar{q}^6 \,,\\
  h_{M}(\bar{q},\beta) &=& \frac{h_{M}^{(0)}(\bar{q},\beta)}{24\beta^4(\beta^2-m^2)^3} + \frac{h_{M}^{(1)}(\bar{q},\beta)}{8\beta^5(\beta^2-m^2)^{\frac{7}{2}}}\,\mathrm{Artanh}\left(\frac{\sqrt{\beta^2-m^2}}{\beta}\right)\,,\\
  h_{M}^{(0)}(\bar{q},\beta) &=& \beta^6(44\beta^4-44\beta^2m^2+15m^4) \nonumber \\ &+& 3\beta^4(24\beta^4-10\beta^2m^2+m^4)\bar{q}^2 + 3\beta^2(12\beta^4+8\beta^2m^2-5m^4)\bar{q}^4 \nonumber \\ &+& (8\beta^4+10\beta^2m^2-3m^4)\bar{q}^6\,,\\
  h_{M}^{(1)}(\bar{q},\beta) &=& \beta^6(-16\beta^6+24\beta^4m^2-18\beta^2m^4+5m^6) \nonumber \\ &-& 3\beta^4m^2(16\beta^4-16\beta^2m^2+5m^4)\bar{q}^2 - \beta^2m^2(32\beta^4-22\beta^2m^2+5m^4)\bar{q}^4 \nonumber \\ &-& m^2(8\beta^4-4\beta^2m^2+m^4)\bar{q}^6\,,\\
  I_{M}(\beta) &=& \frac{\beta^4(8\beta^4+10\beta^2m^2-3m^4)}{192\pi^2(\beta^2-m^2)^3} \nonumber \\
               &-& \frac{3\beta^3(8\beta^4m^2-4\beta^2m^4+m^6)}{192\pi^2(\beta^2-m^2)^{\frac{7}{2}}}\mathrm{Arcosh}\left(\frac{\beta}{m}\right).
\end{eqnarray}
Obviously, the explicit expression for the $I_{B}$ integral can be obtained from $I_{M}$ by replacing $m$ by $M$. Finally, 
in terms of these basic scalar loop functions, the Green function integral $\mathcal{G}(\slashed{p},\beta)$ is expressed as
\begin{equation}
\mathcal{G}(\slashed{p},\beta) = \frac{\slashed{p}}{2s}\left(2\sqrt{s}E_{B}I_{MB}(s,\beta)+I_{M}(\beta)-I_{B}(\beta)\right) + MI_{MB}(s,\beta)\,.
\end{equation}
We also remind the reader that we are working in the c.m.~frame, where $\slashed{p}=\sqrt{s}\gamma^{0}$.

\bibliography{mojeCitace}

\begin{thebibliography}{45}
\expandafter\ifx\csname natexlab\endcsname\relax\def\natexlab#1{#1}\fi
\providecommand{\url}[1]{\texttt{#1}}
\providecommand{\href}[2]{#2}
\providecommand{\path}[1]{#1}
\providecommand{\DOIprefix}{doi:}
\providecommand{\ArXivprefix}{arXiv:}
\providecommand{\URLprefix}{URL: }
\providecommand{\Pubmedprefix}{pmid:}
\providecommand{\doi}[1]{\href{http://dx.doi.org/#1}{\path{#1}}}
\providecommand{\Pubmed}[1]{\href{pmid:#1}{\path{#1}}}
\providecommand{\bibinfo}[2]{#2}
\ifx\xfnm\relax \def\xfnm[#1]{\unskip,\space#1}\fi
\bibitem[{R\'{e}vai(2018)}]{Revai:2017isg}
\bibinfo{author}{J.~R\'{e}vai}, \bibinfo{journal}{Few Body Syst.}
  \bibinfo{volume}{59} (\bibinfo{year}{2018}) \bibinfo{pages}{49}.
\bibitem[{Kaiser et~al.(1995)Kaiser, Siegel, and Weise}]{Kaiser:1995eg}
\bibinfo{author}{N.~Kaiser}, \bibinfo{author}{P.~B. Siegel},
  \bibinfo{author}{W.~Weise}, \bibinfo{journal}{Nucl. Phys.}
  \bibinfo{volume}{A594} (\bibinfo{year}{1995}) \bibinfo{pages}{325--345}.
\bibitem[{Oset and Ramos(1998)}]{Oset:1997it}
\bibinfo{author}{E.~Oset}, \bibinfo{author}{A.~Ramos}, \bibinfo{journal}{Nucl.
  Phys.} \bibinfo{volume}{A635} (\bibinfo{year}{1998})
  \bibinfo{pages}{99--120}.
\bibitem[{Ikeda et~al.(2012)Ikeda, Hyodo, and Weise}]{Ikeda:2012au}
\bibinfo{author}{Y.~Ikeda}, \bibinfo{author}{T.~Hyodo},
  \bibinfo{author}{W.~Weise}, \bibinfo{journal}{Nucl. Phys.}
  \bibinfo{volume}{A881} (\bibinfo{year}{2012}) \bibinfo{pages}{98--114}.
\bibitem[{Ciepl\'{y} and Smejkal(2012)}]{Cieply:2011nq}
\bibinfo{author}{A.~Ciepl\'{y}}, \bibinfo{author}{J.~Smejkal},
  \bibinfo{journal}{Nucl. Phys.} \bibinfo{volume}{A881} (\bibinfo{year}{2012})
  \bibinfo{pages}{115--126}.
\bibitem[{Guo and Oller(2013)}]{Guo:2012vv}
\bibinfo{author}{Z.-H. Guo}, \bibinfo{author}{J.~A. Oller},
  \bibinfo{journal}{Phys. Rev.} \bibinfo{volume}{C87} (\bibinfo{year}{2013})
  \bibinfo{pages}{035202}.
\bibitem[{Mai and Mei{\ss}ner(2015)}]{Mai:2014xna}
\bibinfo{author}{M.~Mai}, \bibinfo{author}{U.-G. Mei{\ss}ner},
  \bibinfo{journal}{Eur. Phys. J.} \bibinfo{volume}{A51} (\bibinfo{year}{2015})
  \bibinfo{pages}{30}.
\bibitem[{Feijoo et~al.(2019)Feijoo, Magas, and Ramos}]{Feijoo:2018den}
\bibinfo{author}{A.~Feijoo}, \bibinfo{author}{V.~Magas},
  \bibinfo{author}{A.~Ramos}, \bibinfo{journal}{Phys. Rev.}
  \bibinfo{volume}{C99} (\bibinfo{year}{2019}) \bibinfo{pages}{035211}.
\bibitem[{Gasser et~al.(1988)Gasser, Sainio, and \v{S}varc}]{Gasser:1987rb}
\bibinfo{author}{J.~Gasser}, \bibinfo{author}{M.~E. Sainio},
  \bibinfo{author}{A.~\v{S}varc}, \bibinfo{journal}{Nucl. Phys.}
  \bibinfo{volume}{B307} (\bibinfo{year}{1988}) \bibinfo{pages}{779--853}.
\bibitem[{Krause(1990)}]{Krause:1990xc}
\bibinfo{author}{A.~Krause}, \bibinfo{journal}{Helv. Phys. Acta}
  \bibinfo{volume}{63} (\bibinfo{year}{1990}) \bibinfo{pages}{3--70}.
\bibitem[{Bernard(2008)}]{Bernard:2007zu}
\bibinfo{author}{V.~Bernard}, \bibinfo{journal}{Prog. Part. Nucl. Phys.}
  \bibinfo{volume}{60} (\bibinfo{year}{2008}) \bibinfo{pages}{82--160}.
\bibitem[{Kaiser(2001)}]{Kaiser:2001hr}
\bibinfo{author}{N.~Kaiser}, \bibinfo{journal}{Phys. Rev.}
  \bibinfo{volume}{C64} (\bibinfo{year}{2001}) \bibinfo{pages}{045204}.
  \bibinfo{note}{[Erratum: Phys. Rev. C73, 069902 (2006)]}.
\bibitem[{Oller and Mei{\ss}ner(2001)}]{Oller:2000fj}
\bibinfo{author}{J.~A. Oller}, \bibinfo{author}{U.-G. Mei{\ss}ner},
  \bibinfo{journal}{Phys. Lett.} \bibinfo{volume}{B500} (\bibinfo{year}{2001})
  \bibinfo{pages}{263--272}.
\bibitem[{Jido et~al.(2003)Jido, Oller, Oset, Ramos, and
  Mei{\ss}ner}]{Jido:2003cb}
\bibinfo{author}{D.~Jido}, \bibinfo{author}{J.~A. Oller},
  \bibinfo{author}{E.~Oset}, \bibinfo{author}{A.~Ramos}, \bibinfo{author}{U.-G.
  Mei{\ss}ner}, \bibinfo{journal}{Nucl. Phys.} \bibinfo{volume}{A725}
  (\bibinfo{year}{2003}) \bibinfo{pages}{181--200}.
\bibitem[{Hyodo and Jido(2012)}]{Hyodo:2011ur}
\bibinfo{author}{T.~Hyodo}, \bibinfo{author}{D.~Jido}, \bibinfo{journal}{Prog.
  Part. Nucl. Phys.} \bibinfo{volume}{67} (\bibinfo{year}{2012})
  \bibinfo{pages}{55--98}.
\bibitem[{Ciepl\'{y} et~al.(2016)Ciepl\'{y}, Mai, Mei{\ss}ner, and
  Smejkal}]{Cieply:2016jby}
\bibinfo{author}{A.~Ciepl\'{y}}, \bibinfo{author}{M.~Mai},
  \bibinfo{author}{U.-G. Mei{\ss}ner}, \bibinfo{author}{J.~Smejkal},
  \bibinfo{journal}{Nucl. Phys.} \bibinfo{volume}{A954} (\bibinfo{year}{2016})
  \bibinfo{pages}{17--40}.
\bibitem[{Shevchenko(2017)}]{Shevchenko:2016wnu}
\bibinfo{author}{N.~V. Shevchenko}, \bibinfo{journal}{Few Body Syst.}
  \bibinfo{volume}{58} (\bibinfo{year}{2017}) \bibinfo{pages}{6}.
\bibitem[{Ohnishi et~al.(2017)Ohnishi, Horiuchi, Hoshino, Miyahara, and
  Hyodo}]{Ohnishi:2017uni}
\bibinfo{author}{S.~Ohnishi}, \bibinfo{author}{W.~Horiuchi},
  \bibinfo{author}{T.~Hoshino}, \bibinfo{author}{K.~Miyahara},
  \bibinfo{author}{T.~Hyodo}, \bibinfo{journal}{Phys. Rev.}
  \bibinfo{volume}{C95} (\bibinfo{year}{2017}) \bibinfo{pages}{065202}.
\bibitem[{Friedman and Gal(2007)}]{Friedman:2007zza}
\bibinfo{author}{E.~Friedman}, \bibinfo{author}{A.~Gal},
  \bibinfo{journal}{Phys. Rept.} \bibinfo{volume}{452} (\bibinfo{year}{2007})
  \bibinfo{pages}{89--153}.
\bibitem[{R\'{e}vai(2019)}]{Revai:2019ipq}
\bibinfo{author}{J.~R\'{e}vai}, \bibinfo{journal}{arXiv:1908.08730[nucl-th]}
  (\bibinfo{year}{2019}).
\bibitem[{Chew et~al.(1957)Chew, Goldberger, Low, and Nambu}]{Chew:1957zz}
\bibinfo{author}{G.~F. Chew}, \bibinfo{author}{M.~L. Goldberger},
  \bibinfo{author}{F.~E. Low}, \bibinfo{author}{Y.~Nambu},
  \bibinfo{journal}{Phys. Rev.} \bibinfo{volume}{106} (\bibinfo{year}{1957})
  \bibinfo{pages}{1337--1344}.
\bibitem[{Liu and Zhu(2007)}]{Liu:2006xja}
\bibinfo{author}{Y.-R. Liu}, \bibinfo{author}{S.-L. Zhu},
  \bibinfo{journal}{Phys. Rev.} \bibinfo{volume}{D75} (\bibinfo{year}{2007})
  \bibinfo{pages}{034003}.
\bibitem[{Mai et~al.(2009)Mai, Bruns, Kubis, and Mei{\ss}ner}]{Mai:2009ce}
\bibinfo{author}{M.~Mai}, \bibinfo{author}{P.~C. Bruns},
  \bibinfo{author}{B.~Kubis}, \bibinfo{author}{U.-G. Mei{\ss}ner},
  \bibinfo{journal}{Phys. Rev.} \bibinfo{volume}{D80} (\bibinfo{year}{2009})
  \bibinfo{pages}{094006}.
\bibitem[{Borasoy et~al.(2007)Borasoy, Bruns, Mei{\ss}ner, and
  Ni{\ss}ler}]{Borasoy:2007ku}
\bibinfo{author}{B.~Borasoy}, \bibinfo{author}{P.~C. Bruns},
  \bibinfo{author}{U.-G. Mei{\ss}ner}, \bibinfo{author}{R.~Ni{\ss}ler},
  \bibinfo{journal}{Eur. Phys. J.} \bibinfo{volume}{A34} (\bibinfo{year}{2007})
  \bibinfo{pages}{161--183}.
\bibitem[{Nieves and Ruiz~Arriola(2001)}]{Nieves:2001wt}
\bibinfo{author}{J.~Nieves}, \bibinfo{author}{E.~Ruiz~Arriola},
  \bibinfo{journal}{Phys. Rev.} \bibinfo{volume}{D64} (\bibinfo{year}{2001})
  \bibinfo{pages}{116008}.
\bibitem[{Lutz and Kolomeitsev(2002)}]{Lutz:2001yb}
\bibinfo{author}{M.~F.~M. Lutz}, \bibinfo{author}{E.~E. Kolomeitsev},
  \bibinfo{journal}{Nucl. Phys.} \bibinfo{volume}{A700} (\bibinfo{year}{2002})
  \bibinfo{pages}{193--308}.
\bibitem[{Djukanovic et~al.(2006)Djukanovic, Gegelia, and
  Scherer}]{Djukanovic:2006xc}
\bibinfo{author}{D.~Djukanovic}, \bibinfo{author}{J.~Gegelia},
  \bibinfo{author}{S.~Scherer}, \bibinfo{journal}{Eur. Phys. J.}
  \bibinfo{volume}{A29} (\bibinfo{year}{2006}) \bibinfo{pages}{337--342}.
\bibitem[{Bruns et~al.(2011)Bruns, Mai, and Mei{\ss}ner}]{Bruns:2010sv}
\bibinfo{author}{P.~C. Bruns}, \bibinfo{author}{M.~Mai}, \bibinfo{author}{U.-G.
  Mei{\ss}ner}, \bibinfo{journal}{Phys. Lett.} \bibinfo{volume}{B697}
  (\bibinfo{year}{2011}) \bibinfo{pages}{254--259}.
\bibitem[{Morimatsu and Yamada(2019)}]{Morimatsu:2019wvk}
\bibinfo{author}{O.~Morimatsu}, \bibinfo{author}{K.~Yamada},
  \bibinfo{journal}{Phys. Rev.} \bibinfo{volume}{C100} (\bibinfo{year}{2019})
  \bibinfo{pages}{025201}.
\bibitem[{Csejthey-Barth et~al.(1965)}]{Csejthey-Barth:1965izu}
\bibinfo{author}{M.~Csejthey-Barth}, et~al., \bibinfo{journal}{Phys. Lett.}
  \bibinfo{volume}{16} (\bibinfo{year}{1965}) \bibinfo{pages}{89--91}.
\bibitem[{Sakitt et~al.(1965)Sakitt, Day, Glasser, Seeman, Friedman, Humphrey,
  and Ross}]{Sakitt:1965kh}
\bibinfo{author}{M.~Sakitt}, \bibinfo{author}{T.~B. Day},
  \bibinfo{author}{R.~G. Glasser}, \bibinfo{author}{N.~Seeman},
  \bibinfo{author}{J.~H. Friedman}, \bibinfo{author}{W.~E. Humphrey},
  \bibinfo{author}{R.~R. Ross}, \bibinfo{journal}{Phys. Rev.}
  \bibinfo{volume}{139} (\bibinfo{year}{1965}) \bibinfo{pages}{B719}.
\bibitem[{Kim(1965)}]{Kim:1965zzd}
\bibinfo{author}{J.~K. Kim}, \bibinfo{journal}{Phys. Rev. Lett.}
  \bibinfo{volume}{14} (\bibinfo{year}{1965}) \bibinfo{pages}{29}.
\bibitem[{Mast et~al.(1976)Mast, Alston-Garnjost, Bangerter, Barbaro-Galtieri,
  Solmitz, and Tripp}]{Mast:1975pv}
\bibinfo{author}{T.~S. Mast}, \bibinfo{author}{M.~Alston-Garnjost},
  \bibinfo{author}{R.~O. Bangerter}, \bibinfo{author}{A.~S. Barbaro-Galtieri},
  \bibinfo{author}{F.~T. Solmitz}, \bibinfo{author}{R.~D. Tripp},
  \bibinfo{journal}{Phys. Rev.} \bibinfo{volume}{D14} (\bibinfo{year}{1976})
  \bibinfo{pages}{13}.
\bibitem[{Bangerter et~al.(1981)Bangerter, Alston-Garnjost, Barbaro-Galtieri,
  Mast, Solmitz, and Tripp}]{Bangerter:1980px}
\bibinfo{author}{R.~O. Bangerter}, \bibinfo{author}{M.~Alston-Garnjost},
  \bibinfo{author}{A.~Barbaro-Galtieri}, \bibinfo{author}{T.~S. Mast},
  \bibinfo{author}{F.~T. Solmitz}, \bibinfo{author}{R.~D. Tripp},
  \bibinfo{journal}{Phys. Rev.} \bibinfo{volume}{D23} (\bibinfo{year}{1981})
  \bibinfo{pages}{1484}.
\bibitem[{Ciborowski et~al.(1982)}]{Ciborowski:1982et}
\bibinfo{author}{J.~Ciborowski}, et~al., \bibinfo{journal}{J. Phys.}
  \bibinfo{volume}{G8} (\bibinfo{year}{1982}) \bibinfo{pages}{13--32}.
\bibitem[{Nowak et~al.(1978)}]{Nowak:1978au}
\bibinfo{author}{R.~J. Nowak}, et~al., \bibinfo{journal}{Nucl. Phys.}
  \bibinfo{volume}{B139} (\bibinfo{year}{1978}) \bibinfo{pages}{61--71}.
\bibitem[{Tovee et~al.(1971)}]{Tovee:1971ga}
\bibinfo{author}{D.~N. Tovee}, et~al., \bibinfo{journal}{Nucl. Phys.}
  \bibinfo{volume}{B33} (\bibinfo{year}{1971}) \bibinfo{pages}{493--504}.
\bibitem[{Bazzi et~al.(2011)}]{Bazzi:2011zj}
\bibinfo{author}{M.~Bazzi}, et~al. (\bibinfo{collaboration}{SIDDHARTA}),
  \bibinfo{journal}{Phys. Lett.} \bibinfo{volume}{B704} (\bibinfo{year}{2011})
  \bibinfo{pages}{113--117}.
\bibitem[{Rosner et~al.(2018)Rosner, Stone, and Van~de Water}]{Rosner:2018pdg}
\bibinfo{author}{J.~L. Rosner}, \bibinfo{author}{S.~Stone},
  \bibinfo{author}{R.~S. Van~de Water} (\bibinfo{collaboration}{Particle Data
  Group}), \bibinfo{journal}{Phys. Rev.} \bibinfo{volume}{D98}
  (\bibinfo{year}{2018}) \bibinfo{pages}{030700}.
\bibitem[{Aoki et~al.(2019)}]{Aoki:2019cca}
\bibinfo{author}{S.~Aoki}, et~al. (\bibinfo{collaboration}{Flavour Lattice
  Averaging Group}), \bibinfo{journal}{arXiv:1902.08191[hep-lat]}
  (\bibinfo{year}{2019}).
\bibitem[{Tanabashi et~al.(2018)}]{Tanabashi:2018oca}
\bibinfo{author}{M.~Tanabashi}, et~al. (\bibinfo{collaboration}{Particle Data
  Group}), \bibinfo{journal}{Phys. Rev.} \bibinfo{volume}{D98}
  (\bibinfo{year}{2018}) \bibinfo{pages}{030001}.
\bibitem[{Mai and Mei{\ss}ner(2013)}]{Mai:2012dt}
\bibinfo{author}{M.~Mai}, \bibinfo{author}{U.-G. Mei{\ss}ner},
  \bibinfo{journal}{Nucl. Phys.} \bibinfo{volume}{A900} (\bibinfo{year}{2013})
  \bibinfo{pages}{51 -- 64}.
\bibitem[{Moriya et~al.(2013)}]{Moriya:2013eb}
\bibinfo{author}{K.~Moriya}, et~al. (\bibinfo{collaboration}{CLAS}),
  \bibinfo{journal}{Phys. Rev.} \bibinfo{volume}{C87} (\bibinfo{year}{2013})
  \bibinfo{pages}{035206}.
\bibitem[{Roca and Oset(2013)}]{Roca:2013av}
\bibinfo{author}{L.~Roca}, \bibinfo{author}{E.~Oset}, \bibinfo{journal}{Phys.
  Rev.} \bibinfo{volume}{C87} (\bibinfo{year}{2013}) \bibinfo{pages}{055201}.
\bibitem[{Nakamura and Jido(2014)}]{Nakamura:2013boa}
\bibinfo{author}{S.~X. Nakamura}, \bibinfo{author}{D.~Jido},
  \bibinfo{journal}{PTEP} \bibinfo{volume}{2014} (\bibinfo{year}{2014})
  \bibinfo{pages}{023D01}.

\end{thebibliography}



\end{document}